\documentclass[twocolumn,showpacs,preprintnumbers,amsmath,amssymb,amsfonts,
superscriptaddress,
nofootinbib,
 aps,
 prb,
floatfix,
lengthcheck,%
]{revtex4-2}

\usepackage[english]{babel} 
\usepackage[latin1]{inputenc}
\usepackage[usenames,dvipsnames]{color}
\usepackage{graphicx}
\usepackage{amsmath,amssymb,amsfonts}
\usepackage[colorlinks=true,linkcolor=blue,urlcolor=blue,citecolor=blue]{hyperref}
\usepackage[percent]{overpic}

\graphicspath{{Figures/}}

\newcommand {\sign} {\mathrm{sign}}

\begin{document}

\title{Correlated two-Leviton states in the fractional quantum Hall regime}

\author{Bruno Bertin-Johannet}
\affiliation{Aix Marseille Univ, Universit\'e de Toulon, CNRS, CPT, Marseille, France}
\email{bruno.bertin@etu.univ-amu.fr}
\author{Alexandre Popoff}
\affiliation{Aix Marseille Univ, Universit\'e de Toulon, CNRS, CPT, Marseille, France}
\affiliation{Coll\`ege de Tipaerui, BP 4557 - 98713 Papeete, Tahiti, French Polynesia}
\author{Flavio Ronetti}
\affiliation{Aix Marseille Univ, Universit\'e de Toulon, CNRS, CPT, Marseille, France}
\author{J\'er\^ome Rech}
\affiliation{Aix Marseille Univ, Universit\'e de Toulon, CNRS, CPT, Marseille, France}
\author{Thibaut Jonckheere}
\affiliation{Aix Marseille Univ, Universit\'e de Toulon, CNRS, CPT, Marseille, France}
\author{Laurent Raymond}
\affiliation{Aix Marseille Univ, Universit\'e de Toulon, CNRS, CPT, Marseille, France}
\author{Benoit Gr\'emaud}
\affiliation{Aix Marseille Univ, Universit\'e de Toulon, CNRS, CPT, Marseille, France}
\author{Thierry Martin}
\affiliation{Aix Marseille Univ, Universit\'e de Toulon, CNRS, CPT, Marseille, France}

\begin{abstract}
We consider a two-dimensional electron system in the Laughlin sequence of the fractional quantum Hall regime to investigate the effect of strong correlations on the mutual interaction between two Levitons, single-electron excitations generated by trains of quantized Lorentzian pulses. We focus on two-Leviton states injected in a single period with a time separation $\Delta t$. In the presence of a quantum point contact operating in the weak-backscattering regime, we compute the backscattered charge by means of the Keldysh technique. In the limit of an infinite period and zero temperature, we show that the backscattered charge for a two-Leviton state is not equal to twice the backscattered charge for a single Leviton. We present an interpretation for this result in terms of  the wave-packet formalism for Levitons, thus proposing that an effective interaction between the two Levitons is induced by the strongly-correlated background. Finally, we perform numerical calculations in the periodic case by using the Floquet formalism for photo-assisted transport. By varying the system parameters such as pulse width, filling factor and temperature we show that the value of the backscattered charge for two-Leviton states is strongly dependent on the pulse separation, thus opening scenarios where the effective interaction between Levitons can be controllably tuned. 
\end{abstract}

\maketitle

\section{Introduction}
The manipulation of individual quantum systems is at the heart of current research in physics aiming at fostering the development of new applications in the domain of quantum technologies~\cite{raimond2001,haroche2013,blais2021}. Historically, the principal effort has been devoted to the generation and control of photons, the quanta of light, leading to the proposal of quantum computation scheme based on single-photon states~\cite{knill2001,kok2007,peruzzo2014,lodahl2018,wehner2018}. Recently, a fast progress in nanoelectronics  paved the way towards the manipulation of single-electron states~\cite{glattli2016b,bauerle2018,edlbauer2022}. This great interest in the generation and control of single-electron states has lead to the development of a new research field which has been called Electron Quantum Optics (EQO)~\cite{degiovanni09,bocquillon12,bocquillon14}.

The major step at the foundations of EQO has been the experimental realization of single-electron sources~\cite{bauerle2018}. The first proposal to inject a single electron into the filled Fermi sea of a mesoscopic channel was introduced by B\"uttiker and collaborators and is known as the mesoscopic capacitor~\cite{buttiker1993a,buttiker1993b}. In their proposal, the energy levels of a quantum dot are periodically driven leading to the alternate emission of an electron and a hole into a two-dimensional electron system~\cite{ferraro14,ferraro2015}. Despite the fact that this source has been actually realized in experiments~\cite{gabelli06,feve07}, it requires complex nanolitography techniques to properly design quantum dots and it is also difficult to imagine how to generalize this scheme to the emission of multiple electronic excitations in the same period. Both issues can be overcome by considering a different injection scheme based on a time-dependent drive~\cite{levitov96,ivanov97}. As shown by Levitov and co-workers~\cite{keeling06}, a train of quantized Lorentzian voltage pulses injects single- and few-electron states, namely propagating wave-packets carrying a single electron devoid of additional particle-hole pairs, into ballistic quantum channels. These minimal excitations have been called Levitons and they have been proved to induce the minimal excess current noise when injected into ballistic channels of meso-scale devices. Due to the intrinsic property of this source, $q$ different Levitons can be injected in a single period and travel unhindered along ballistic channels. The many-body states that are consequently formed are called multi-electron Levitons or, simply, $q$-Leviton states~\cite{glattli16,ronetti2018crystallization,moskalets2018}.

The tools of quantum transport have been widely employed in the context of EQO to investigate the properties of these excitations propagating in ballistic edge channels. In particular, the electrical shot noise~\cite{martin05}, induced by the granularity of electrons, has proven to be an invaluable source of information to probe the discrete nature of propagating single-electron states. Ground-breaking experimental results have opened the way to the triggered emission and manipulation of single-electron excitations by adapting quantum optics results to the realm of condensed matter. In the first phase of EQO these experiments have shown that it is possible to reproduce the phenomenology of standard quantum optics in mesoscopic fermionic systems in the absence of interaction between electrons  \cite{bocquillon12,dubois13a}, by replacing the bosonic statistics of photons with the Fermi-Dirac statistics. Interestingly, compared to photons, electrons can interact with each other and with the electromagnetic background, thus rendering extremely appealing to investigate the effects of the electron-electron interaction on single-electron states. Indeed, a great interest has been devoted to the study of the effects of different types of solid-state correlations on the propagation of Levitons, for instance in systems with superconductivity~\cite{acciai2019a,ronetti2020,bertin-johannet2022} or Coulomb interaction~\cite{acciai2018,ferraro2018,vannucci2018,ronetti2019,acciai2019b,rebora2020,rebora2021a}.
 
 The fractional quantum Hall effect (FQHE) represents a seminal example of strongly-correlated states where the interaction between electrons cannot be neglected~\cite{tsui82,laughlin83}. In the Laughlin sequence of the FQHE a single chiral channel exists at the boundary of the system and the excitations are exotic quasi-particles with fractional charge and statistics called anyons~\cite{nayak08,hashisaka2021,jonckheere2022,glidic2023}. The propagation of Levitons in these exotic states of matter is currently under investigation. Importantly, it has been shown that quantized Lorentzian pulses still injects minimal single-electron excitations even in fractional quantum Hall channels~\cite{rech17}. Intriguingly, the propagation of Levitons in these edge states has proven to lead to non-trivial properties, which have no counterparts for non-interacting systems~\cite{ronetti2018crystallization}. Indeed, in the case of multi-Levitons, the strongly-correlated background re-arranges the charge density into an oscillating pattern after a tunneling at the QPC, leading to the so-called crystallization of Levitons, in analogy with the formation of Wigner crystals in one dimensional systems~\cite{wigner1934,schulz1993,deshpande2010}. These works prove that the injection of  Levitons in a fractional quantum Hall bar still presents many non-trivial aspects to investigate. 

 Recently, in the context of EQO, the idea of implementing quantum information and computation schemes based on the concept of electron flying qubit has attracted a lot of attention~\cite{dasenbrook2015,dasenbrook2016,dasenbrook2016b,glattli2016b,bauerle2018,edlbauer2022}. In analogy with previous proposals for photonic states, the qubit states are defined by the presence or the absence of an electron in two alternative propagation paths, which are termed quantum rails. Even in the absence of electron-electron interaction, single-qubit gates can be realized by extended tunneling regions coupling two or more quantum rails in the presence of voltage gates~\cite{bauerle2018,edlbauer2022}. The angle of rotation on the Bloch sphere is proportional to the length of the tunneling region and the voltage applied to the gates, thus allowing for a full control on the single-qubit operation.  In contrast, the presence of an interaction between electron flying qubits is a crucial ingredient towards the realization of two-qubit quantum gates and, therefore, universal quantum computation schemes in coherent semiconducting nano-electronics systems~\cite{takeda2017,edlbauer2022}. The Coulomb interaction introduces a quantum phase between the two states which is at the origin of the entanglement required to implement a two-qubit gate. However, so far no theoretical proposal or experimental evidence has ever revealed the interaction between two propagating Levitons. 
 
 In this paper, we intend to elucidate this matter and propose a way to measure the effect of an induced interaction between Levitons. Our main emphasis is on the detection of the effect of this interaction between Levitons and not of the proposal of quantum information schemes based on the aforementioned interaction. For this purpose, we employ a two-dimensional electron system tuned into the Laughlin sequence of the FQHE in a four-terminal configuration to measure charge transport properties. We focus on periodic trains of two-Leviton states separated by a delay $\Delta t$, i.e. we set the parameter of the Lorentzian voltage to inject two Levitons per period. As proven in Ref.~\cite{vannucci2017}, multiple Levitons are still minimal electronic excitations even in the presence of a finite time delay. By considering a QPC operating in the weak-backscattering regime, we compute the backscattered charge as a function of the separation time between Levitons. We first focus on the limit of a single pulse (i.e., infinite period), thus showing that the backscattered charge for a two-Leviton state, namely $Q_2$ is not equal to twice the backscattered charge $Q_1$ for a single Leviton. We re-derive the same result  by resorting to the wave-packet formalism for Levitons, showing that $Q_2$ contains a term with four Leviton wave-functions, in contrast to $Q_1$ which is expressed in terms of an overlap of only two wave-functions. We intepret this additional term in $Q_2$ as a result of an effective interaction between the two Levitons induced by the strongly-correlated background. Then, we consider the periodic case and we perform a numerical calculation to show that $Q_2$ differs from twice $Q_1$ even for a finite period. Indeed, we show that the time delay $\Delta t$ between the two pulses can be exploited to reduce, remove or increase the interaction between Levitons in the periodic case. Finally we address the case of the separated injection of a Leviton and an anti-Leviton and show that the backscattered charge is independent of the time delay. We conjecture that our results are valid in any type of Luttinger liquid, including for instance the integer quantum Hall effect at $\nu = 2$. The choice of focusing on the Laughlin sequence is motivated by the absence of decoherence induced by other propagating channels~\cite{bocquillon13,wahl14,marguerite2016}. 

The paper is organized as follows. In Sec.~\ref{sec:Model}, we introduce the model of the quantum Hall bar with quantum point contact and the source of Levitons. In Sec.~\ref{sec:Charge}, we compute the backscattered charge at the quantum point contact for isolated pulses of Levitons and for a periodic drive: we interpret our results by using the wave-packet formalism for Levitons. In Sec.~\ref{sec:Results}, we discuss our results by plotting the backscattered charge as a function of different system parameters. In Sec.~\ref{sec:Conclusions}, we draw our conclusions.

%
%
\begin{figure}
	\includegraphics[width=\linewidth]{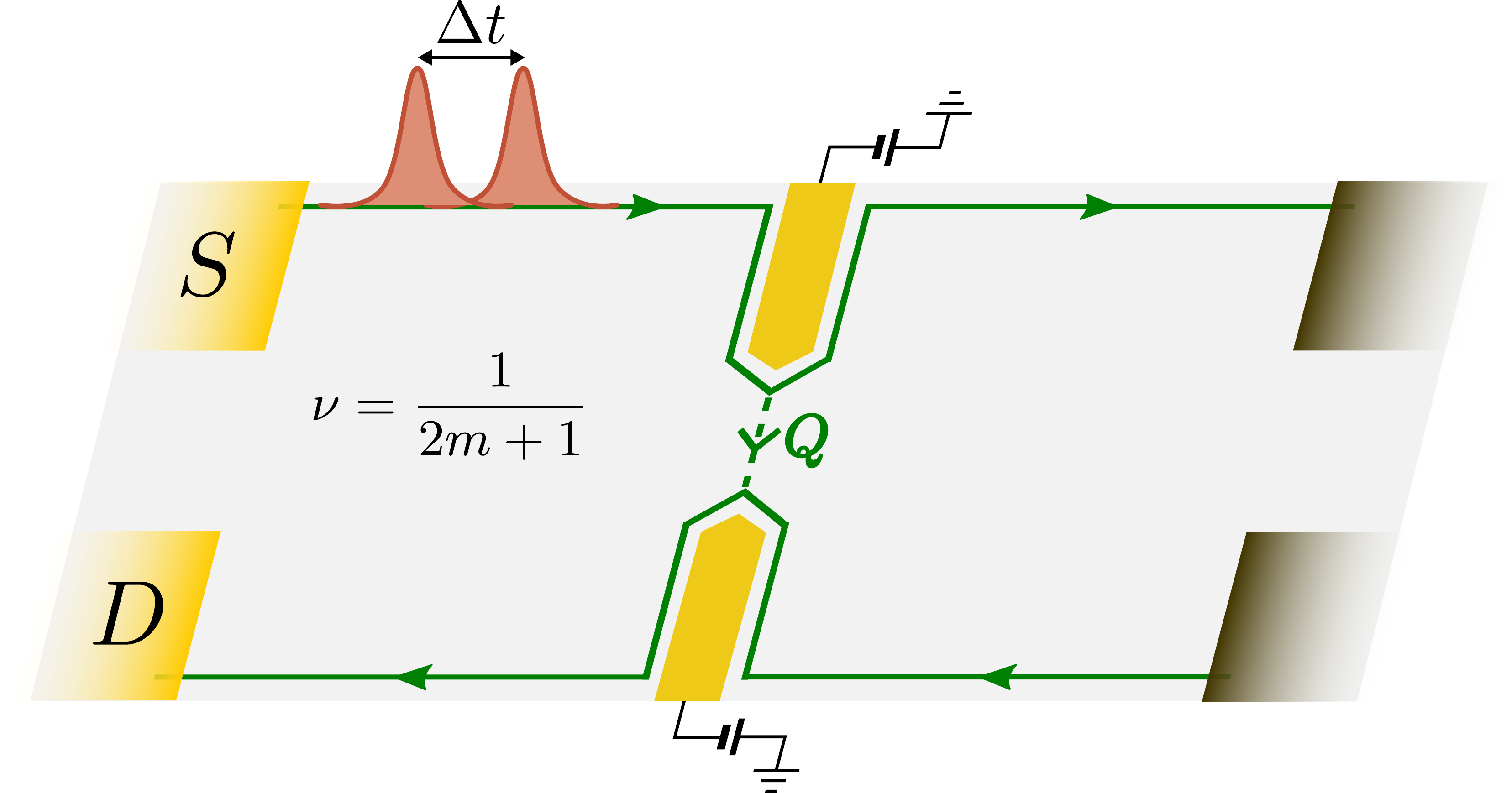}
	\caption{(Color online) Schematic view of the setup. A two-dimensional electron system is in the Laughlin series of the FQHE, with a single edge state propagating on the boundary. The system is connected to four terminals: a periodic voltage drive applied to the source terminal $S$ is injecting multiple Levitons in each period separated by a time delay $\Delta t$ smaller than the period itself. The oppposite edges are connected by a QPC placed in $x=0$. The charge backscattered at the QPC, namely $Q$, is measured in the drain $D$. The remaining terminals (in grey) are grounded and they are not involved in any measurement.}
	\label{fig:setup}
\end{figure}
\section{Model \label{sec:Model}} We consider a FQHE bar at filling factor $\nu = 1/(2n+1)$, with $n \in \mathbb{N}$: at this filling factor only one chiral edge mode emerges at the boundary of the sample (see Fig.~\ref{fig:setup}). Since the bulk of the system is insulating, transport of electrons occurs only along the chiral modes on the two opposite edges. These modes are connected by a QPC. The latter operates in the weak-backscattering regime, where the tunneling is dominated by Laughlin quasi-particles with charge $e^* = \nu e$~\cite{kane92}. The FQHE bar is connected to four reservoirs. A time-dependent voltage $V(t)$ is applied to terminal $S$ and the charge backscattered at the QPC is measured in the drain $D$. The two remaining terminals are assumed to be grounded.  The total Hamiltonian that describes this system is given by $H = H_{\rm 0} + H_{\rm s} + H_{\rm B}$ and consists of edge states, time-dependent drive and tunneling terms respectively. By using chiral Luttinger theory \cite{Wen95}, the effective Hamiltonian for the edge states reads ($\hbar=1$)
\begin{equation}
	\label{eq:H_wg}
	H_{\rm 0} = \sum_{r=R,L} \frac{v}{4\pi} \int \mathrm{d}x\, \left[ \partial_x \Phi_r(x) \right]^2.
\end{equation}
The right and left moving excitations chirally propagating along the two edges are described in terms of bosonic fields $\Phi_{R/L}$, satisfying $[\Phi_{R/L}(x), \Phi_{R/L}(y)] = \pm i \pi \sign(x-y)$ and $v$ is their velocity. The charge density is defined in terms of bosonic fields as
\begin{align}
\rho_{R/L} (x) = \pm \frac{e \sqrt{\nu}}{2\pi} \partial_x \Phi_{R/L}(x).
\end{align}
The driving voltage is thus capacitively coupled to the density of right-moving excitations by means of the following term in the Hamiltonian
\begin{equation}
\label{H_s}
	H_{\rm s} = V(t)\int \mathrm{d}x \, \Theta(x-d)  \rho_R(x).
\end{equation}
Here, the step function $\Theta(x-d)$ describes an infinite and homogeneous contact to which the time-dependent voltage $V(t)$ is applied. Equations of motion for the bosonic field $\Phi_{R}$ in the presence of the source term are solved in terms of the single-variable fields $\phi_{R}$ in the equilibrium configuration $V=0$, thus giving (see for instance Ref~\cite{vannucci2017})
\begin{equation}
\label{eq:bosofields}
	\Phi_{R}\left(t-\frac{x}{v}\right) = \phi_{R}\left(t -\frac x v \right) - \frac{1}{\sqrt \nu} \varphi\left(t-\frac{x}{v_F}\right).
\end{equation}
where $\varphi(t)$ is the phase difference between the two electrodes and is defined as
\begin{equation}
\varphi(t)=e^*\int_{-\infty}^{t}\mathrm{d}t'V(t').
\end{equation}
The propagation of bosonic excitations along the edge states remains chiral even in the presence of the driving voltage: this is a consequence of the linear dispersion of edge states for all filling factors in the Laughlin sequence. The profile of these excitations is determined by the time-dependence of the phase difference $\varphi(t)$. Therefore, the choice of the driving potential is crucial to define the nature of the propagating modes. Here, we focus on periodic trains of Lorentzian-shaped voltage pulses. We consider quantized pulses carrying an integer charge $-qe=\frac{e^2\nu}{2\pi}\int_{0}^{\mathcal{T}}\mathrm{d}t\, V(t)$, where $q$ is any integer number, here named $q$-Levitons. In order to study the effect of the background correlations on $q$-Levitons we take into account the possibility of injecting multiple Levitons in one period separated by a delay $\Delta t$. The corresponding time-dependent potential is
\begin{equation}
	\label{eq:Levitons}
V(t)=\sum_{j=0}^{q-1}\sum\limits_{k=-\infty}^{+\infty}\frac{V_0}{\pi}\frac{\gamma^2}{\gamma^2+(t-k \mathcal{T}-j \Delta t)^2},
\end{equation}
with period $\mathcal{T}=\frac{2\pi}{\omega}$, amplitude $V_0$ and width $2\gamma$.  Later, we will consider also the case of an isolated pulse, which can be recovered from the above expression in the limit $\gamma \ll \mathcal{T}$.

Finally, we consider the tunneling between the two edges which occurs through a QPC at $x=0$. For this reason we can  set $x=0$ in Eq.~\eqref{eq:bosofields} and consider only the time-dependence of fields. Assuming that the QPC is working in the weak backscattering regime, the tunneling of Laughlin quasiparticles between opposite edges is the only relevant process \cite{Kane94,Saminadayar97,dePicciotto97}. Annihilation fields for Laughlin quasiparticles carrying fractional charge $-\nu e$ (with $e>0$) are defined through the standard procedure of bosonization \cite{Wen95}. They read 
\begin{equation}
\Psi_{R/L}(t) = \frac{\mathcal{F}_{R/L}}{\sqrt{2\pi a}} e^{-i \sqrt \nu \Phi_{R/L}(t)} = e^{i \varphi(t)}\psi_{R/L}(t),
\end{equation}
where $a$ is a short-distance cut-off and $\mathcal{F}_{R/L}$ are the Klein factors \cite{Wen95,vondelft98,guyon2002,martin05} and we introduced the fermionic fields at equilibrium ($V=0$)
\begin{equation}
	\psi_{R/L}(t) = \frac{\mathcal{F}_{R/L}}{\sqrt{2\pi a}} e^{-i \sqrt \nu \phi_{R/L}(t)}.
\end{equation}
The backscattered Hamiltonian reads
\begin{equation}
H_{\rm B}=\lambda\sum_{\epsilon=+,-}\left[\Psi_R^\dagger(t)\Psi_L(t)\right]^\epsilon\, ,
\end{equation}
where $\epsilon=-$ means Hermitian conjugate. By expressing the Hamiltonian in terms of fields at equilibrium, we obtain
\begin{equation}
H_{\rm B}=\lambda\sum_{\epsilon=+,-}\left[e^{-i\varphi(t)} \psi_R^\dagger(t)\psi_L(t)\right]^\epsilon .
\end{equation}
\section{Charge  backscattered at the QPC\label{sec:Charge}} 
We are interested in investigating the properties of the charge backscattered at the QPC. Indeed, the transport properties before the QPC does not carry any information about the effect of strong correlations of the propagation of $q$-Levitons, as shown in Ref.~\cite{ronetti2018crystallization}. Nevertheless, we expect that, due  to the non-linear nature of the tunneling at the QPC,  effects of interaction will be manifested in the backscattered charge. For a periodic time-dependent voltage, the charge which is backscattered in one period $\mathcal{T}$ is given by
\begin{align}
\label{eq:QB}
Q = \int_{-\mathcal{T}/2}^{\mathcal{T}/2} \mathrm{d}t\, \left\langle I_B(t) \right\rangle\, ,
\end{align}
where we introduce the backscattering current
\begin{equation}
I_{\rm B}(t)=ie^*\lambda\sum_{\epsilon=+,-}\epsilon\left[e^{i\varphi(t)}\psi_R(t)^\dagger \psi_L(t)\right]^\epsilon\, .
\end{equation}
The assumption of weak backscattering regime allows us to calculate the excess charge density perturbatively in the tunneling Hamiltonian $H_{\rm B}$. Thermal averages are thus performed over the initial equilibrium density matrix in the absence of tunneling. In order to properly manage the out-of-equilibrium dynamics of the system, calculations are usefully carried out in the Keldysh formalism. To lowest order in the tunneling amplitude $\lambda$, the backscattered charge becomes
\begin{equation}\label{eq:charge2int} 
\begin{aligned}
	Q=i\frac{e^*\lambda^2}{2\pi^2a^2}&\int_{-\infty}^{+\infty}\mathrm{d}\tau\, e^{2\nu \mathcal{G}\left(\tau\right)}\\
	&\times\int_{-\mathcal{T}/2}^{\mathcal{T}/2}\mathrm{d}t\,\sin\left[\varphi(t)-\varphi(t-\tau)\right],
\end{aligned}
\end{equation}
where we introduced the connected bosonic Green's function $\mathcal{G}\left(\tau\right)=\left\langle \phi_{R/L}(\tau) \phi_{R/L}(0)\right\rangle_c$. Its expression for a finite temperature $\theta$ reads ($k_{\rm B } = 1$)
\begin{align}
\mathcal{G}\left(\tau\right) = \log \left[\frac{\pi \theta \tau}{\sinh\left(\pi \theta \tau\right)\left(1+i \frac{\tau }{\tau_0}\right)}\right],
\end{align}
with $\tau_0 = a/v_F$ is the short-time cut-off: our theoretical description is valid for times much longer than $\tau_0$.

The expressions for the backscattered charge $Q$ is valid for any arbitrary driving voltage $V(t)$. In the  following, we focus specifically on the case of $q$-Levitons, by using the time-dependent potential defined in Eq.~\eqref{eq:Levitons} for $q = 1$ and $q = 2$. In the latter case, we will assume a time delay $\Delta t$ between the two Levitons. Before numerically evaluating $Q$ for a finite period $\mathcal{T}$, we start by considering the case of isolated pulses ($\gamma \ll \mathcal{T}$) and zero temperature. In this case, we can provide analytical expressions for the backscattered charge that will set the stage for our later discussion. 

For the sake of completeness, we will also present the calculations for the charge fluctuations in the case of isolated pulses at zero temperature. This would provide a full characterization of the transport properties of this setup in the presence of one-Leviton and two-Levitons states.

\subsection{Isolated pulses}
\subsubsection{Backscattered charge \label{sec:ChargePulse}}
In the case of isolated pulses the integral over $t$ in Eq.~\eqref{eq:QB} can be extended from $-\infty$ to $+\infty$ and can be solved analytically. Let us comment that this limit is well-defined only for voltage pulses that go to zero at $t=\pm \infty$, which is the case for Lorentzian-shaped pulses. The expression for the charge becomes
\begin{equation}\label{eq:charge3int} 
\begin{aligned}
	Q=i\frac{e^*\lambda^2}{2\pi^2a^2}&\int_{-\infty}^{+\infty}\, \mathrm{d}\tau e^{2\nu \mathcal{G}(\tau)}\\
	&\times\int_{-\infty}^{+\infty}\mathrm{d}t\,\sin\left[\varphi(t)-\varphi(t-\tau)\right].
\end{aligned}
\end{equation}
The integral over $t$ can be solved analytically for integer values of $q$. For $q = 1$, one finds
\begin{equation}
	Q_1=\frac{4 ie^*}{\pi a^2}\lambda^2\gamma^2\int_{-\infty}^{+\infty}\mathrm{d}\tau\, e^{2\nu \mathcal{G}(\tau)}\frac{\tau}{\tau^2+4\gamma^2}\, .
\end{equation}
Next, we consider the case $q=2$ where the isolated pulses are separated by a constant delay $\Delta t$. The integral over $t$ gives
\begin{widetext}
	\begin{equation}\label{eq:totalchargetransfer}
	Q_2=\frac{16ie^*}{\pi a^2}\lambda^2\gamma^2\int_{-\infty}^{+\infty}\mathrm{d}\tau\,e^{2\nu \mathcal{G}(\tau)}\frac{\tau \left[\left(4 \gamma^2 +  \Delta t^2\right)^2 - \tau^2\left(3  \Delta t^2 + 4\gamma^2\right) + 2 \tau^4 \right]}{(\tau^2+4\gamma^2)\left[( \Delta t+\tau)^2+4\gamma^2\right]\left[( \Delta t-\tau)^2+4\gamma^2\right]}\, .
	\end{equation}
\end{widetext}
These two expressions can be computed numerically at finite temperature $\theta$. Nevertheless, one can obtain an analytically expression for $Q_1$ and $Q_2$. Indeed, by using zero-temperature limit of the bosonic Green's function
\begin{align}
	\mathcal{G}\left(\tau\right) = -\log \left(1+i \frac{\tau}{\tau_0}\right)\, ,
\end{align}
the residue theorem can be used to calculate the integral over $t$, thus arriving to the final expression for the backscattered charge in the case $q=1$,
\begin{equation}
 Q_1=e^*\left(\frac{\lambda}{v_F}\right)^2\left(\frac{2\gamma}{\tau_0}\right)^{2-2\nu}\,  + \mathcal{O}\left[\left(\frac{\tau_0}{\gamma}\right)^{2\nu-1}\right],
\end{equation}
where we kept only the leading order in $\tau_0/\gamma$. By performing a similar calculation for $Q_2$ in the limit of zero temperature, we find at the lowest order in $\tau_0/\gamma$
\begin{equation}\label{eq:charge2lev}
\begin{aligned}
 Q_2&=Q_1\Bigg\{2\Re\left[\left(1+\frac{2i\gamma}{ \Delta t}\right)^2\left(1-\frac{i \Delta t}{2\gamma}\right)^{-2\nu}\right] \\
 &+2\left(1+\frac{4\gamma^2}{ \Delta t^2}\right) \Bigg\} + \mathcal{O} \left[ \left(\frac{\tau_0}{\gamma}\right)^{2\nu-1} \right].
\end{aligned}
\end{equation}
\begin{figure}
	\includegraphics{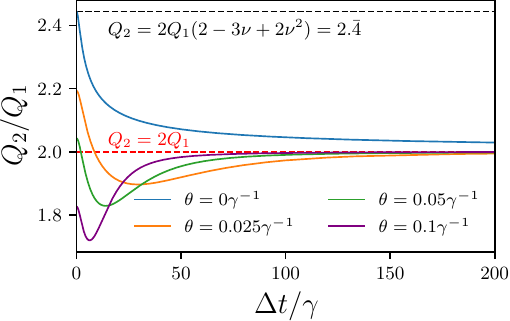}
	\caption{(Color online) Backscattered charge for a two-Leviton state $Q_2$ rescaled with respect to the same quantity for a single Leviton as a function of $\Delta t/\gamma$ at different temperatures. The blue line corresponds to the analytical formula at zero temperature and valid in the limit $\tau_0\ll \gamma$. Black and red dotted lines are respectively the limits of simultaneous ($\Delta t/\gamma \rightarrow 0$) and well-separated pulses ($\Delta t/\gamma \rightarrow \infty$). The other curves have been computed numerically by fixing $\tau_0 = 10^{-4}\gamma$. The only other parameter is the filling factor $\nu = 1/3$.}
	\label{fig:analytical}
\end{figure}
We note that, at zero temperature, the backscattered charge for two pulses is proportional to the backscattered charge for a single pulse. For the integer filling factor $\nu = 1$, we recover the trivial result that $Q_2 = 2 Q_1$. 
Nevertheless, for fractional filling factor, the constant of proportionality depends on $\Delta t$ and $\gamma$ and $Q_2\ne 2Q_1$. 
From this result we can conclude that the charge backscattered when two Levitons are impinging at the QPC does not amount in general to twice the charge backscattered when a single Leviton is injected. 
This result is a consequence of the strong correlations that characterize Laughlin states and that introduce a non-linear current-to-voltage characteristic in the presence of the tunneling of quasi-particles at the QPC .\\
Before concluding this part, it is instructive to analyze two extreme limits of the ratio $ \Delta t/\gamma$ at zero temperature. In the limit of simultaneous pulses, which can be obtained by setting $ \Delta t/\gamma\ll1$ in Eq.~(\ref{eq:charge2lev}), we find
\begin{equation}\label{eq:charge2levsim}
\begin{aligned}
\lim_{\Delta t/\gamma\rightarrow 0}Q_2=2Q_1\left(2 - 3\nu + 2\nu^2\right)\, .
\end{aligned}
\end{equation}
In this limit the constant of proportionality acquires a simple expression, becoming independent of $\gamma$ and being determined solely by the filling factor $\nu$.\\
Finally, we consider the opposite case of well-separated pulses, which can be found by taking the limit $ \Delta t/\gamma\to\infty$ in the expression for $Q_2$,
\begin{equation}
\lim_{\Delta t/\gamma\rightarrow \infty}Q_2 = 2Q_1.
\end{equation}
Indeed, in this case, we recover for any filling factor the trivial result that the charge backscattered for two Levitons is twice the one obtained with a single Leviton. For well-separated injection time, the system has relaxed to equilibrium when the second pulse comes in. As a result, the two Levitons behave as two independent single pulses. 

We summarize these results in Fig.~\ref{fig:analytical}, where we also present some curves computed numerically at finite temperature. Here, the black and red dotted lines represents the two limits of simultaneous and well-separated pulses at zero temperature. Indeed, we note that the limit of well-separated pulses is hard to reach due to the long-tail nature of Lorentzian pulses. It is interesting to point out that, when the temperature is increased, the ratio shows a minimum which gets closer and closer to $0$, whose position is proportional to the inverse temperature $\theta^{-1}$.

\subsubsection{Charge fluctuations}
In order to further characterize the transport properties of this system, we provide here the analytical calculations of charge fluctuations, which are defined as 
\begin{equation}
\mathcal{S}_Q =\int_{-\infty}^{+\infty} dt\int_{-\infty}^{+\infty} dt' \left\langle I_B(t) I_B(t')\right\rangle - Q^2.
\end{equation}
Similarly to the backscattered charge, we compute the above integral at zero temperature and for isolated quantized Lorentzian pulses, in order to provide the analytical expressions of charge fluctuations for $q=1$ and $q=2$. Before providing the results, we notice that charge fluctuations diverges at zero temperature: as discussed in Ref.~\cite{Jonckheere05}, the integral should be regularized by subtracting its value at equilibrium, i.e in the absence of voltage pulses. Therefore, one finds
\begin{align}\label{eq:chargefluct}
	\tilde{\mathcal{S}}_Q=\left(\frac{e^*\lambda}{\pi a}\right)^2&\int_{-\infty}^{+\infty}\mathrm{d}\tau\,e^{2\nu \mathcal{G}\left(\tau\right)}\nonumber\\&\times \int_{-\infty}^{\infty}\mathrm{d}t\, \left\{\cos\left[\varphi(t)-\varphi(t-\tau)\right]-1\right\}\, ,
\end{align}
where we choose the notation $\tilde{\mathcal{S}}_Q$ for the regularized charge fluctuations. The calculation of the integrals over $t$ can be performed analytically for integer values of $q$ using the residue theorem, thus obtaining
\begin{equation}\label{eq:fluctinter1}
\tilde{\mathcal{S}}_{Q,1}=-e^*\frac{4e^*\lambda^2\gamma}{\pi v_\text{F}^2\tau_0^2}\int_{-\infty}^{+\infty}\mathrm{d}\tau\, \frac{\tau^2}{\tau^2+4\gamma^2}e^{2\nu \mathcal{G}\left(\tau\right)}\, 
\end{equation}
for $q=1$ and
\begin{widetext}
\begin{equation}
    	\tilde{\mathcal{S}}_{Q,2}=-2\pi\gamma\left(\frac{2e^*\lambda}{\pi a}\right)^2\int_{-\infty}^{\infty}\mathrm{d}\tau \tau^2\frac{32\gamma^4 + \left(\Delta t^2 - \tau^2\right)^2 + 4\gamma^2\left(3\Delta t^2 - \tau^2\right)}{\Delta t\left(\tau^2+4\gamma^2\right)\left[\left(\Delta t+\tau\right)^2+4\gamma^2\right]\left[\left(\Delta t-\tau\right)^2+4\gamma^2\right]}\left(1+i\frac{\tau}{\tau_0}\right)^{-2\nu}\,
\end{equation}
\end{widetext}
for $q=2$. The residue theorem can be used once again to perform both integrals over $\tau$ and the final expressions for the charge fluctuations for $q=1$ and $q=2$ are
\begin{equation}
	\tilde{\mathcal{S}}_{Q,1}=2\left(\frac{2e^*\lambda\gamma}{v_F\tau_0}\right)^2\left(1+\frac{2\gamma}{\tau_0}\right)^{-2\nu}\, .
\end{equation}
and
\begin{widetext}
\begin{equation}
\label{eq:fluct2levfin}
\tilde{\mathcal{S}}_{Q,2}=2\left(\frac{e^*\lambda}{v_\text{F}}\right)^2\left(\frac{2\gamma}{\tau_0}\right)^{2-2\nu}\Bigg[2\left(1+\frac{4\gamma^2}{\Delta t^2}\right) +2\Re \left(1+\frac{2i\gamma}{\Delta t}\right)^2\left(1-\frac{i \Delta t}{2\gamma}\right)^{-2\nu}\Bigg]\, ,
\end{equation}
\end{widetext}
where we performed the limit $\tau_0 / \gamma \rightarrow 0$. It is instructive to introduce a Fano factor for the charge as 
\begin{equation}
        F_{Q}=\frac{\tilde{\mathcal{S}}_{Q,{1,2}}}{Q_{1,2}}\, .
\end{equation}
We notice that $F_{Q}=2e^*$ for $q=1$ and also for $q=2$ independently of the separation between the Levitons. It is important to remark that the above Fano factor has been defined by employing the regularized charge fluctuations in order to yield Fano relation similar to those existing between current and noise in other setups, see e.g. Refs.~\cite{rech2017a} and \cite{vannucci2017}. Due to the proportionality between regularized charge fluctuations and backscattered charge, the plot of $\mathcal{S}_{Q,{2}}/\mathcal{S}_{Q,{1}}$ as a function of $\Delta t/\gamma$ has the exact same behaviour as the one depicted in Fig.~\ref{fig:analytical}.

Before showing the numerical results for the periodic case, we present a physical interpretation of the above results in terms of the wave-packet formalism of Levitons.
\subsection{Correlated two-Levitons state}
In this part, we recast the expression for the backscattered charges $Q_1$ and $Q_2$ at zero temperature in terms of the wave-function of an isolated Leviton. Indeed, a quantized Lorentzian drive with a single peak creates a quantum state of the form~\cite{keeling06,Grenier13}
\begin{equation}
\left|\Psi\right\rangle = \int dx ~\mathcal{X}^*(x)\psi^{\dagger}(x)\left|F\right\rangle,
\end{equation}
where $\psi^{\dagger}(x)$ creates an electron at the position $x$, $\left|F\right\rangle$ is the ground state of the system and
\begin{align}
	\label{eq:wave-leviton}
	\mathcal{X}(x-vt) = \sqrt{\frac{\gamma v}{\pi}}\frac{1}{x-vt+i v\gamma},
\end{align}
is the wave-function of a single Leviton propagating in a chiral edge state. For the following discussion, it is useful to introduce the excess charge $\Delta Q$ as 
\begin{equation}
	\label{eq:deltaQ}
	\Delta  Q \equiv Q_2 -2 Q_1,
\end{equation}
which represents the backscattered charge in excess compared to the trivial case at $\nu = 1$. By using Eq.~\eqref{eq:wave-leviton}, the charge $Q_1$ can be recast as
\begin{equation}
\begin{aligned}
	Q_1 = -\frac{e^*\lambda^2}{2\pi^2 a^2} \int_{-\infty}^{\infty}\mathrm{d}t\, \int_{-\infty}^{\infty}\mathrm{d}\tau\,  &\Re\left[\chi(t)\chi^*(t-\tau)\right]\\
	\times&\tau \left(1+i\frac{\tau}{\tau_0}\right)^{-2\nu}\, ,
\end{aligned}
\end{equation}
where we defined $\chi(t) \equiv \mathcal{X}(-vt)$. We observe that the charge $Q_1$ contains a product of Leviton wave-function $\chi$, thus showing that it is determined directly by the charge density of the state injected on the system ground state. One can similarly express the excess charge $\Delta Q$ in terms of Leviton wave-functions, thus obtaining
\begin{equation}
\begin{aligned}
	\Delta Q =\frac{e^*\lambda^2}{4\pi^2 a^2(i)^{2\nu-2}}\int_{-\infty}^{\infty}\mathrm{d}t \big[&\chi(t)g_{\nu}(t,\Delta t)\chi(t+\Delta t)\\
	&\qquad -\text{h.c.}\big]
\end{aligned}
\end{equation}
where 
\begin{equation}
	g_{\nu}(t,\Delta t) = \int_{-\infty}^{\infty}\mathrm{d}\tau\, \chi^*(t-\tau)\chi^*(t-\tau+\Delta t)\tau^{2-2\nu}.
\end{equation}
In contrast with $Q_1$, the excess charge is related to the product of four Leviton wave-functions, thus proving that it is originated by a density-density interaction between Levitons. We interpret this result by conjecturing that the strongly-correlated background mediates an effective interaction between the two Levitons. In the above expression we set $\tau_0 = 0$ since the integrated function is well-behaved around $\tau=0$. Let us observe that for $\nu = 1$ (no interactions), this function vanishes
\begin{align}
	g_{\nu = 1}(t,\Delta t) = \int_{-\infty}^{\infty}d\tau \chi^*(t-\tau)\chi^*(t-\tau+\Delta t) = 0.
\end{align}
The function $g_{\nu}$ is different from zero for fractional filling factors because of the propagator $e^{2\nu \mathcal{G}(\tau)}$. The power law decay for fractional filling factors is slower than $\tau^2$, thus inducing long-time correlations between the two Levitons. These correlations do not affect the charge only when the pulses are well separated (limit of $\Delta t$ very big). Otherwise, correlations induce an effective interaction between Levitons that effectively enhance the value of the charge $Q_2$ compared to the limit of two well-isolated pulses.


\subsection{Periodic train of Levitons}
In order to make contact with experiments, we consider here the periodic case since the emission of a single isolated pulse is still experimentally challenging. We will show that not only our results still hold but also that new phenomena emerge. In the case of a periodic signal, no analytical expression can be derived for the backscattered charges $Q_1$ and $Q_2$ and one has to resort to a numerical calculation. The latter is conveniently carried on in the photo-assisted formalism, where the transport properties are expressed in terms of the Fourier coefficients of the phase associated with the periodic signal~\cite{crepieux2006}. In the following, we will focus only on the case of Lorentzian-shaped pulses with $q=1$ and $q=2$.\\The photo-assisted expressions for the backscattered charges $Q_1$ and $Q_2$ are
\begin{align}
	\label{eq:Q1}
	Q_1 &=\mathcal{Q} \sum_{m}p_m^2\left\lvert\Gamma\left(\nu+i\frac{m+1}{2\theta\pi}\right)\right\rvert^2\sinh\left(\frac{m+1}{2\theta}\right),\\ \label{eq:Q2}
	Q_2 &=\mathcal{Q} \sum_{m}\tilde{p}_m^2\left\lvert\Gamma\left(\nu+i\frac{m+2}{2\theta\pi}\right)\right\rvert^2\sinh\left(\frac{m+2}{2\theta}\right),
\end{align}
where $\mathcal{Q} = \frac{2e^*}{\mathcal{T}}\left(\frac{\lambda}{ v}\right)^2\left(2\pi\theta \tau_0\right)^{2\nu-2}\frac{\theta}{\Gamma(2\nu)}$. Here, we introduced the photo-assisted coefficients for $q=1$
\begin{equation}
	\label{eq:pm}
p_m = \begin{cases}
		e^{-2\pi \eta m}\left(1-e^{-4\pi \eta}\right)\hspace{5mm} &m\ge 0\\
		-e^{-2\pi \eta}\hspace{5mm} &m = -1\\
		0\hspace{5mm} &m< -1.
	\end{cases}
\end{equation}
and for $q=2$
\begin{equation}
	\tilde{p}_m = \begin{cases}
		\frac{\left[1-e^{i \pi \alpha (m+1)}\right]-e^{-4\pi \eta -i \pi \alpha}\left[1-e^{i \pi \alpha (m+3)}\right]}{\left(1-e^{i \pi \alpha}\right)e^{i \pi \alpha m}}p_m &m\ge 0\\e^{i \pi \alpha}\left(e^{-i \pi \alpha}+1\right)\left(1-e^{-4\pi \eta}\right) p_{-1} & m=-1 \\ e^{i \pi \alpha} p_{-1}^2.&m=-2\\0 &m<-2
	\end{cases}.
\end{equation}
in terms of the rescaled pulse width $\eta = \gamma / \mathcal{T}$, the reduced temperature $\theta = k_{\rm B} T /\hbar\omega$ and the pulse separation $\alpha = 2\Delta t/\mathcal{T}$. The sums appearing in Eqs.~\eqref{eq:Q1} and~\eqref{eq:Q2} can be evaluated numerically: their convergence is assured by the negative exponential of coefficients $p_m$ in Eq.~\eqref{eq:pm}. As for the case of isolated pulses, we can define an excess charge $\Delta Q = Q_2 - 2Q_1$ for the periodic drive. We present the corresponding results in the next section by carrying on an analysis over different system parameters. Moreover, we checked that the interpretation of an interaction between Levitons induced by the background can be generalized to the periodic case by resorting to the periodic version of Leviton's wave-functions~\cite{glattli16, ronetti2018crystallization}.

Concerning the charge fluctuations in the periodic case, in Ref.~\cite{vannucci2017} some of the authors have already shown that a Fano relation exist between noise and backscattered current for one-Leviton and multi-Leviton states regardless of the value of the time separation $\Delta t$. Therefore, one can conclude that, given the proportionality between charge fluctuations and backscattered charge, the calculation of the latter quantity is sufficient to fully characterize the transport properties of the system. Moreover, it is instructive to point out that the regularized charge fluctuations introduced for isolated pulse in Eq.~\eqref{eq:chargefluct} is the correct expression that is recovered from the periodic case in the limit $\eta \to 0$.

Before concluding this section, we address the case of the injection of two excitations with opposite charge, i.e. one Leviton and one anti-Leviton, separated by a time delay $\Delta t$ in the same period. The form of the applied voltage is
\begin{equation}
	\label{eq:LevitonsAntiLevitons}
	V(t)=\sum\limits_{s=\pm }s\sum\limits_{k=-\infty}^{+\infty}\frac{V_0}{\pi}\frac{\gamma^2}{\gamma^2+(t-k \mathcal{T}+s \Delta t/2)^2}.
\end{equation}
We note that the voltage is an even function of time, i.e $V(t) = V(-t)$. By using this property, in Appendix~\ref{app:A}, we show that the backscattered charge induced by this voltage vanishes for any value of $\Delta t$. We concluded that for the drive in Eq.~\eqref{eq:LevitonsAntiLevitons}, the total backscattered charge $Q_2$ is always identically zero for any choice of system parameters.

The interaction between the Leviton and the anti-Leviton cannot be probed by a measure of the backscattered charge.
A computation of the charge fluctuations can therefore prove to be useful.
To this end, we performed the equivalent of the calculation leading to Eq.~(\ref{eq:fluct2levfin}).
This allowed to obtain the charge fluctuations induced by a single couple of Leviton anti-Leviton, which read
\begin{equation}
        \tilde{S}_{Q,l\overline{l}} = 2\left[\left(\frac{2\gamma}{\Delta t}\right)^2+1\right]^{-1} \tilde{S}_{Q,l}\, .
\end{equation}
Where $l$ stands for Leviton and $\overline{l}$ stands for anti-Leviton.
This result can be interpreted as follows, when $\Delta t/\gamma\to0$, the Leviton is superposed with the anti-Leviton in the drive, resulting in no drive at all, and when $\Delta t/\gamma\to\infty$ they are well separated and give twice the fluctuations of a single Leviton.

Finally, it is worth pointing out that the case of Leviton and anti-Leviton is equivalent to an HOM interferometer, where two Levitons with the same charge are injected from two opposite terminals, e.g. terminal $S$ and the gray terminal of left moving particles in our setup Fig.~\ref{fig:setup}, separated by a controllable time delay. Indeed, we notice that the charge fluctuations for the case of two excitations with opposite charge separated by a time delay correspond to the well-known analytical expression for the charge noise computed in an Hong-Ou-Mandel configuration, see for instance Ref.~\cite{rech2017a}.

\section{Results \label{sec:Results}}
In the previous section we established a connection between the excess charge $\Delta Q$ and the effect of the strongly-correlated background on the two-Leviton state. Here, we discuss our results and present some plots of the latter quantity as a function of different parameters. 
\subsection{Comparison with the isolated pulse case and effects of time separation}
\begin{figure}
	\includegraphics[width=\linewidth]{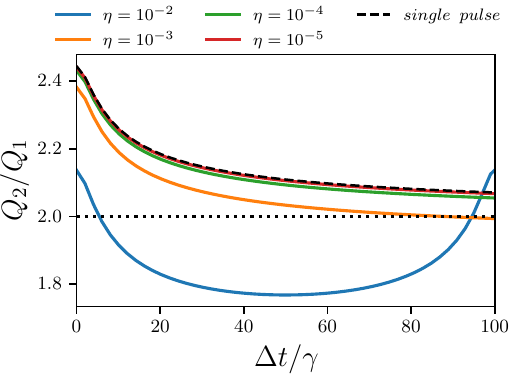}
	\caption{(Color online) Backscattered charge for a two-Leviton state $Q_2$ rescaled with respect to the same quantity for a single Leviton as a function of $\Delta t/\gamma$ at zero temperature. The black dashed line is the limit of a single pulse. Solid lines are computed for the periodic case with $\eta = 10^{-2}, 10^{-3}, 10^{-4}, 10^{-5}$. We see that for the smallest value of $\eta = 10^{-5}$, the periodic case coincides with the analytical limit at infinite period.  The black dotted is a visual guide for $Q_2 = 2Q_1$. The ratio $Q_2 / Q_1$ always stays above this line, except for the highest value of $\eta = 10^{-2}$: in this case $Q_2$ can be smaller than twice $Q_1$.  The only other parameter is the filling factor $\nu = 1/3$.}
	\label{fig:fig3}
\end{figure}

Before investigating the excess charge, in Fig.~\ref{fig:fig3}, we plot the backscattered charge $Q_2$ at filling factor $\nu = 1/3$, normalized with respect to $Q_1$, as a function of $\Delta t / \gamma$ in the periodic case for different values of the parameter $\eta = \gamma / \mathcal{T}$. 
Here, in order to provide an estimation for the experimentally realistic value of $\gamma$, we focus on Lorentzian pulses tailored for future applications in the quantum information domain. 
In order to couple long enough single-electron excitations in the micrometric tunneling region, short pulses should be realized. 
It has been estimated that the required pulse width to perform a single-qubit rotation is roughly $\gamma\sim 10$ ps, which is at the limit of state-of-the-art technology~\cite{edlbauer2022}. 
Shorter pulses can be envisaged by resorting to schemes based on optoelectronics generation~\cite{heshmat2012}. 
Since in usual experiments with Levitons, the frequency is set to $\Omega = 2\pi \times 5$ Ghz, the resulting value for the renormalized width is $\eta = 5 \cdot 10^{-2}$. 
While in the following analysis we will consider values of $\eta$ close to this estimation, in Fig.~\ref{fig:fig3} we consider also smaller values of $\eta$ in order to compare the charge backscatterd with a periodic drive to the same quantity for the case of isolated pulses. 
One can clearly see that for $\eta = 10^{-4}, 10^{-5}$ the colored lines corresponding to the periodic case approach the black dashed line, which corresponds to the case of isolated pulses. This is in agreement with the idea that the case of isolated pulses is recovered  in the limit $\gamma \ll \mathcal{T}$, as anticipated in the model section. 
From the analysis of Fig.~\ref{fig:fig3}, we can also gather some information concerning a different qualitative behavior of $Q_2$ with respect to the case of isolated pulses. Indeed, we could check that for $\eta = 10^{-4}, 10^{-5}$ the ratio $Q_2 / Q_1$ is always greater than or equal to $2$:  the effect of the correlated background is always to enhance the backscattered charge compared to the case $\nu = 1$ as for the case of an isolated pulse. Indeed, for $\eta = 10^{-2}, 10^{-3}$, we see that the ratio $Q_2/Q_1$ could be also smaller or equal to $2$, thus showing that the effect of the correlated background is strongly affected by the width of Lorentzian pulses in the periodic case. In passing, we comment that this additional feature appears exactly for values of $\eta$ which are closer to realistic estimation.
\begin{figure}
	\includegraphics[width=\linewidth]{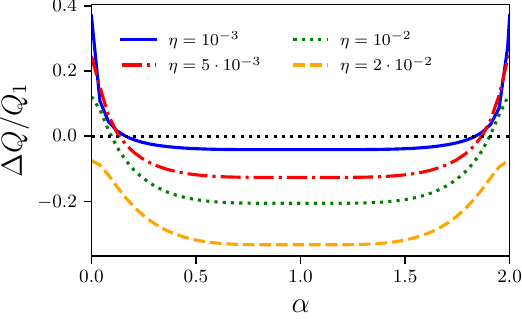}
	\caption{(Color online) Excess backscattered charge $\Delta Q$ rescaled with respect to the backscatterd charge $Q_1$ for a single Leviton as a function of $\alpha = 2\Delta  t/ \mathcal{T}$ at zero temperature. The black dotted line is a visual guide for $\Delta Q = 0$. Solid lines are computed for the periodic case with $\eta = 10^{-3}, 5\cdot 10^{-3}, 10^{-2}, 2\cdot 10^{-2}$. The excess charge $\Delta Q$ changes sign as a function of $\alpha$, except for the highest value of $\eta = 2\cdot 10^{-2}$: in this case $\Delta Q$ is always negative. The smaller the value of $\eta$ and the higher the value of $\alpha$ at which $\Delta Q = 0$. The only other parameter is the filling factor $\nu = 1/3$.}
	\label{fig:fig4}
\end{figure}For this reason, we explore further the dependence of the backscattered charge on parameters $\alpha = 2\Delta t/\mathcal{T}$ and $\eta = \gamma / \mathcal{T}$. In particular, we focus on the quantity $\Delta Q$ introduced in Eq.~\eqref{eq:deltaQ}: for $\Delta Q>0$ ($\Delta Q<0$), the backscattered charge is increased (reduced) by strong correlations with respect to the trivial case at $\nu = 1$. In Fig.~\ref{fig:fig4}, we present the excess charge as a function of $\alpha$ for different values of $\eta$. One can deduce from this plot that, for a large range of $\eta$, the sign of $\Delta Q$ can be changed by tuning the parameter $\alpha$.
\begin{figure}
	\includegraphics[width=\linewidth]{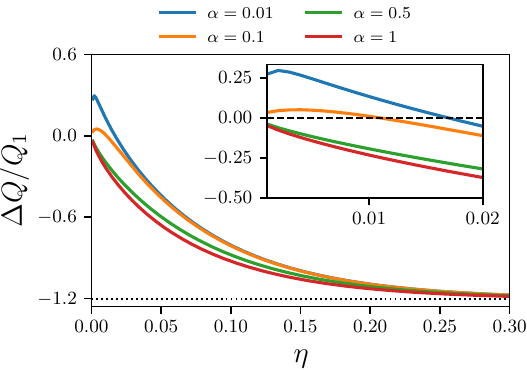}
	\caption{(Color online) Excess backscattered charge $\Delta Q$ rescaled with respect to the backscatterd charge $Q_1$ for a single Leviton as a function of $\eta = \gamma/ \mathcal{T}$ at zero temperature. The black dotted is a visual guide for the limit $\eta \rightarrow \infty$: in this case the periodic drive corresponds to a constant voltage with $q = 2$ and one finds the analytical value of $\Delta Q = Q_1 \left(2^{2\nu -1} -2\right)$. The other lines are computed for the periodic case with $\alpha = 0.01, 0.1, 0.5, 1$. We see that for increasing values of $\eta$, all lines tend to the same limit $\Delta Q = Q_1 \left(2^{2\nu -1} -2\right)$, independently of $\alpha$. The only other parameter is the filling factor $\nu = 1/3$. (Inset) Zoom of the same plot between $\eta = 0$ and $\eta = 2 \cdot 10^{-2}$: one can see that $\Delta Q$ changes sign only for $\alpha = 0.01$ and $\alpha = 0.1$.}
	\label{fig:fig5}
\end{figure}
Interestingly, there exist some values of $\alpha$ where the excess charge $\Delta Q$ vanishes, thus showing that the effect of strong correlations on two-Leviton states can be tuned on and off by acting on the separation time $\Delta t$. Above a certain value of $\eta$, we found that the sign of $\Delta Q$ is negative for any value of $\alpha$ at zero temperature and $\nu = 1/3$. While the specific values of $\eta$ depend on temperature and filling factor, the important result is that there always exists a width of Lorentzian pulse above which the sign of $\Delta Q$ is strictly negative. 

\subsection{Effects of the pulse width}
Similarly, in Fig.~\ref{fig:fig5} we explore the behavior of the excess charge as a function of $\eta$. In particular, we observe that for increasing values of $\eta$ all the lines reach the asymptotic excess charge $\Delta Q = Q_1 \left(2^{2\nu -1} -2\right)$, which corresponds to the limit of a constant voltage $V_{DC} = \hbar \Omega q / e$. This is in agreement with the fact that, the case $\eta \rightarrow \infty$ correspond to a constant signal. In the inset of this figure, we zoom in on these curves in the range $\eta = 10^{-3} - 10^{-2}$. Again, we observe that for some values of $\alpha$ and $\eta$ the excess charge $\Delta Q$ vanishes and by tuning these parameters its sign can be reversed. 
\begin{figure}
	\includegraphics[width=\linewidth]{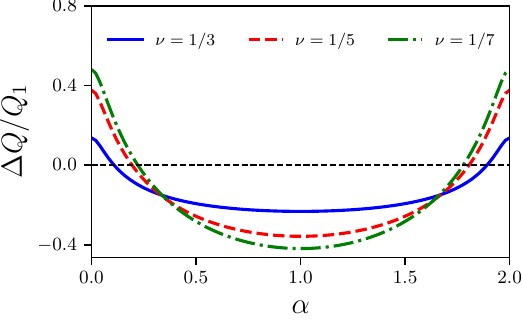}
	\caption{(Color online) Excess backscattered charge $\Delta Q$ rescaled with respect to the backscattered charge $Q_1$ for a single Leviton as a function of $\alpha = 2\Delta  t/ \mathcal{T}$ at zero temperature. The black dotted is a visual guide for $\Delta Q = 0$ and corresponds to the case $\nu = 1$. The other lines are computed for the periodic case with $\nu = 1/3, 1/5, 1/7$. The excess charge $\Delta Q$ is changing sign as a function of $\alpha$ for all the fractional filling factors. The smaller the value of $\nu$, i.e. the stronger the interaction, the higher is the value of $\alpha$ at which $\Delta Q $ changes sign. The only other parameters is $\eta = 10^{-2}$.}
	\label{fig:fig6}
\end{figure}
\subsection{Effects of the filling factor and the temperature}
Finally, we discuss the dependence of $\Delta Q$ from the properties of the strongly-correlated background. The latter are encoded in the Green's function which in turn depends on the filling factor $\nu$ and the temperature $\theta$. The dependence of $\Delta Q$ on the filling factor and, therefore, the strength of the interaction is presented in Fig.~\ref{fig:fig6} for $\nu = 1/3, 1/5, 1/7$. In general, we observe two behaviors as the filling factor is reduced. First of all, the values of $\alpha$ for which the excess charge vanishes increase, thus implying that a stronger interaction requires larger separation time $\Delta t$ to compensate its effect on the two-Leviton state. Moreover, for stronger correlations, the absolute value of the maximum (in $\alpha = 0$) and the minimum (in $\alpha = 1$) of $\Delta Q$ are also increasing. 
\begin{figure}
	\includegraphics[width=\linewidth]{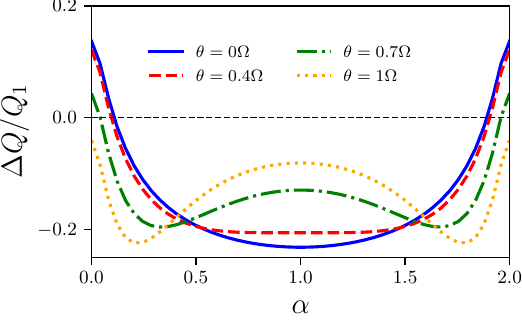}
	\caption{(Color online) Excess backscattered charge $\Delta Q$ rescaled with respect to the backscattered charge $Q_1$ for a single Leviton as a function of $\alpha = 2\Delta  t/ \mathcal{T}$ for different values of temperature. The black dotted line is a visual guide for $\Delta Q = 0$. The other lines are computed for the periodic case with $\theta = 0, 0.4 \Omega, 0.7 \Omega, \Omega$. As the value of temperature is increased, a plateau develops around $\alpha=1$. For higher temperatures, two local minima appear at the left and at the right of $\alpha = 1$.  The excess charge $\Delta Q$ is changing sign as a function of $\alpha$, except for the highest value of $\theta = \Omega$: in this case $\Delta Q$ is always negative. The other parameters are $\nu = 1/3$ and $\eta = 10^{-2}$.}
	\label{fig:fig7}
\end{figure}
We study the effect of temperature in Fig.~\ref{fig:fig7}. Here, we observe a qualitatively different behavior as a function of $\alpha$ for different regimes of temperature. For low temperature, the excess charge behaves as the zero-temperature case (blue line): in this case, a single minimum appears at $\alpha = 1$. As the temperature is increased, a plateau develops for intermediate values of $\alpha$, similarly to the case of isolated pulses presented in Fig.~\ref{fig:analytical}. The width of this plateau is roughly proportional to $\theta^{-1}$. In the high temperature limit, two minima appear for values of $\alpha$ different from $\alpha = 1$ and whose position scales with the inverse temperature, much like the isolated pulse case of Sec.~\ref{sec:ChargePulse}. In this regime, the value of $\Delta Q$ in $\alpha = 1$ corresponds to a local maximum. In order to make contact with experiments, we comment that for $\Omega = 2\pi \times 5$ Ghz, the temperature $\theta = \Omega$ corresponds to roughly  $1.5$ K.

\section{Conclusions \label{sec:Conclusions}}
Here, we consider a quantum Hall bar in the Laughlin sequence of the fractional quantum Hall effect to investigate the effect of a strongly-correlated system on the propagation of two-Leviton states. These states are injected by periodic trains of quantized Lorentizian-shaped pulses with width $\gamma$ and are separated by a controllable $\Delta t$. They are minimal electronic excitations, i.e. purely electronic states travelling above the Fermi sea and generating the minimal electrical noise. \color{red}In \color{black} the presence of a quantum point contact, we compute the backscattered charge for the low-transparency limit. We present analytical results in the case of isolated pulses (i.e., infinite period) and zero temperature. We consider the charge backscattered for the two-Leviton state, namely $Q_2$, and we compare it with the backscattered charge in the presence of a single pulse, termed $Q_1$. By an explicit calculation, we show that $Q_2 > 2 Q_1$ in the fractional regime, in contrast to the trivial result $Q_2 = 2 Q_1$ at $\nu =1$.  Interestingly, in the limit of simultaneous pulses ($\Delta t\ll \gamma$), the backscattered charge aquires a simple expression depending only on the filling factor $\nu$. By resorting to the wave-packet formalism for Levitons, we conjecture the existence of an effective interaction between the two Levitons caused by the strongly-correlated background. Then, we consider the case of a finite period, performing numerical calculations in the photo-assisted formalism. We show that in the periodic case, as a function of $\Delta t$, one can tune the backscattered charge $Q_2$ to be smaller than, equal to or greater than $2 Q_1$. The effect of the correlated background are considered in the dependence of the backscattered charge on the filling factor and on temperature.  We based our numerical calculations on realistic estimations of the parameters that can be realized with state-of-the-art technology. Interesting extension of the present work can be the calculation of the backscattered charge in the presence of extended or multiple quantum point contacts, thus taking into account the effect of quantum interference~\cite{chevallier2010,vannucci2015,ronetti2016}. Indeed, the realization of quantum computing architecture with Levitons requires the presence of multiple tunneling regions.

\begin{acknowledgements}
One of us (TM) thanks A. Lebedev for early discussions on this topic.
This work received support from the French government under the France 2030 investment plan, as part of the Initiative d'Excellence d'Aix-Marseille Universit\'e A*MIDEX. We acknowledge support from the institutes IPhU (AMX-19-IET008) and AMUtech (AMX-19-IET-01X). D.C.G. acknowledges the ANR FullyQuantum 16-CE30-0015-01 grant and the H2020 FET-OPEN UltraFastNano No. 862683 grant.
\end{acknowledgements}
\appendix
 \section{Leviton and anti-Leviton case \label{app:A}}
In this Appendix, we compute the charge backscatterd at the QPC when terminal $1$ is driven by the voltage
\begin{equation}
	V(t)=\sum\limits_{s=\pm }s\sum\limits_{k=-\infty}^{+\infty}\frac{V_0}{\pi}\frac{\gamma^2}{\gamma^2+(t-k \mathcal{T}+s \Delta t/2)^2},
\end{equation}
which corresponds to the injection of a Leviton and an anti-Leviton in the same period separated by a time delay $\Delta t$. In this case, $V(t) = V(-t)$. The phase $\varphi(t) $ becomes in this case
\begin{equation}
	\begin{aligned}
		\varphi(t) &= e^{*}\int_{-\infty}^{t}\mathrm{d}t'\,\sum\limits_{s=\pm }s\sum\limits_{k=-\infty}^{+\infty}\frac{V_0}{\pi}\frac{\gamma^2}{\gamma^2+(t'-k \mathcal{T}+s \Delta t/2)^2}\\
		&=e^{*}\int_{t-\Delta/2}^{t+\Delta/2}\mathrm{d}t'\,\sum\limits_{k=-\infty}^{+\infty}\frac{V_0}{\pi}\frac{\gamma^2}{\gamma^2+(t'-k \mathcal{T})^2},
	\end{aligned}
\end{equation}
and, as a result, $\varphi(t) = \varphi(-t)$. By using the latter property, we show that the integral over $t$ in Eq.~\eqref{eq:charge3int} vanishes. Indeed,
\begin{equation}
	\begin{aligned}
	&\int_{-\infty}^{\infty}\mathrm{d}t\,\sin\left[\varphi(t)-\varphi(t-\tau)\right]\\&=\int_{-\infty}^{\infty}\mathrm{d}t\,\sin\left[\varphi(-t+\tau/2)-\varphi(-t-\tau/2)\right].
\end{aligned}
\end{equation}
By using $\varphi(t) = -\varphi(-t)$, we obtain
\begin{equation}
\begin{aligned}
&\int_{-\infty}^{\infty}\mathrm{d}t\,\sin\left[\varphi(t+\tau/2)-\varphi(t-\tau/2)\right]\\&=\int_{-\infty}^{\infty}\mathrm{d}t\,\sin\left[\varphi(t-\tau/2)-\varphi(t+\tau/2)\right],
\end{aligned}
\end{equation}
and the integral is obviously zero.
\bibliography{ns-levitons.bib}
\end{document}